\documentstyle[11pt,twoside,epsfig]{article}

\newcommand{\La}{\mbox{$\cal L$}}

\newcommand{\zit}{\bibitem}
\newcommand{\proc}[2]{\mbox{$ #1 \rightarrow #2 $}}
\newcommand{\X}{\mbox{\rm hadrons}}
\newcommand{\GammaP}{\mbox{$\gamma p$}}
\newcommand{\ptmiss}{\mbox{$V$}}

\newcommand{\ptnu}{\mbox{$p_{\perp}$}}

\newcommand{\Qsquare}{\mbox{$Q^2$}}
\newcommand{\SignalRegion}{{\em signal region}}
\newcommand{\unit}[1]{{\rm \,#1}}
\newcommand{\GeV}{\unit{GeV}}

\newcommand{\ns}{\unit{ns}}
\newcommand{\ptvec}{\mbox{$\vec{p}_{\perp}$}}

\newcommand{\CC}{\mbox{$CC$}}
\newcommand{\NC}{\mbox{$NC$}}

\newcommand{\Scc}{\mbox{$S$}}
\newcommand{\Vcc}{\mbox{$V$}}
\newlength{\dinwidth}
\newlength{\dinmargin}
\begin{document}
\setlength{\unitlength}{1cm}
\begin{titlepage}
\begin{flushleft}
%
%
{\tt DESY 95-102    \hfill    ISSN 0418-9833} \\
{\tt hep-ex/9506002} \\
{\tt May 1995}                  \\
\end{flushleft}

\vspace*{4.cm}
\begin{center}
\begin{Large}
 {\bf Measurement of the  $e^+$ and $e^-$ induced charged current \\
     cross sections at HERA} \\
\vspace{1.cm}
{H1 Collaboration}    \\
\end{Large}
%
\vspace*{4.cm}
{\bf Abstract:}
\end{center}
\begin{quotation}
\renewcommand{\baselinestretch}{1.0}\large\normalsize
The cross sections for the charged current processes
\proc{e^{-}p}{\nu_e+\X} and, for the first time,
\proc{e^{+}p}{\overline{\nu}_e+\X} are measured at HERA for transverse
momenta larger than 25\GeV.
\renewcommand{\baselinestretch}{1.2}\large\normalsize
\begin{center}
\vfill
%
%
To be submitted to Zeitschrift f\"ur Physik C
\cleardoublepage
\end{center}
\end{quotation}
\end{titlepage}
\begin{Large} \begin{center} H1 Collaboration \end{center} \end{Large}
 S.~Aid$^{13}$,                   
 V.~Andreev$^{25}$,               
 B.~Andrieu$^{28}$,               
 R.-D.~Appuhn$^{11}$,             
 M.~Arpagaus$^{36}$,              
 A.~Babaev$^{24}$,                
 J.~B\"ahr$^{35}$,                
 J.~B\'an$^{17}$,                 
 Y.~Ban$^{27}$,                   
 P.~Baranov$^{25}$
% ...
$ ^{11}$ DESY, Hamburg, Germany$^a$ \\
% ...
$ ^{13}$ II. Institut f\"ur Experimentalphysik, Universit\"at Hamburg,
                                     Hamburg, Germany$^ a$  \\
% ...
$ ^{17}$ Institute of Experimental Physics, Slovak Academy of
                Sciences, Ko\v{s}ice, Slovak Republic$^ f$\\
% ...
$ ^{24}$ Institute for Theoretical and Experimental Physics,
                                                 Moscow, Russia \\
 $ ^{25}$ Lebedev Physical Institute, Moscow, Russia$^ f$ \\
% ...
$ ^{27}$ LAL, Universit\'{e} de Paris-Sud, IN2P3-CNRS,
                            Orsay, France\\
 $ ^{28}$ LPNHE, Ecole Polytechnique, IN2P3-CNRS,
                             Palaiseau, France \\
% ...
$ ^{35}$ DESY, Institut f\"ur Hochenergiephysik,
                              Zeuthen, Germany$^ a$\\
 $ ^{36}$ Institut f\"ur Teilchenphysik,
          ETH, Z\"urich, Switzerland$^ i$\\
% ...
$ ^a$ Supported by the Bundesministerium f\"ur
                                  Forschung und Technologie, FRG
 under contract numbers 6AC17P, 6AC47P, 6DO57I, 6HH17P, 6HH27I, 6HD17I,
 6HD27I, 6KI17P, 6MP17I, and 6WT87P \\
% ...
$ ^f$ Supported by the Deutsche Forschungsgemeinschaft\\
% ...
$ ^i$ Supported by the Swiss National Science Foundation\\


%
\newpage
%
\section{Introduction}
%
With the $ep$-collider HERA the investigation of weak charged currents,
extensively studied in fixed target experiments\cite{hp},
has been resumed at a distance scale at which the propagator of the charged
weak boson $W$ plays a prominent role\cite{cc94,zeus}.
The operation of HERA in 1994 was started with $e^-$ beams and was then
switched to $e^+$ beams allowing considerably increased currents. This opens
the possibility to study simultaneously $W^+$ and $W^-$ induced
processes at very high energies in analogy to low energy neutrino and
antineutrino scattering.

\begin{figure*}[hbtp]\centering
\epsfig{file=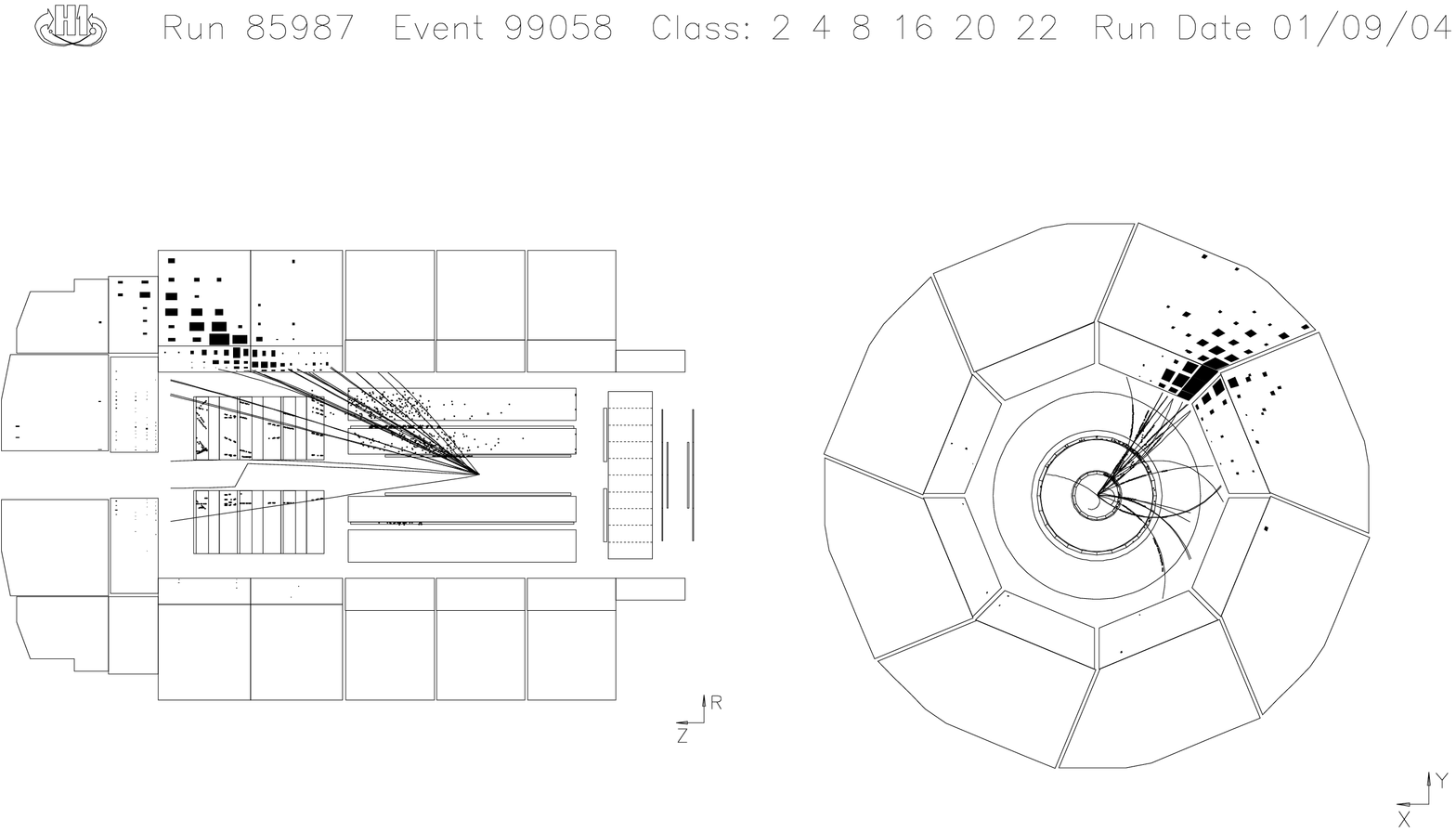,
bbllx=50pt,bblly=40pt,bburx=450pt,bbury=810pt,clip=,angle=90,width=13cm}
\caption{\label{fig:ccevt} \sl A candidate for the process
           \proc{e^+p}{\overline{\nu}_e+\X} with $Q^2$ of about
           20000 GeV$^2$ and $Bjorken-x$ of about 0.4 recorded in the H1
           detector. The views along (the proton beam enters from the right,
           the positron beam from the left) and perpendicular to the beam
           are shown. }.
\end{figure*}

This article presents the first  measurement of the total cross section
\proc{e^{+}p}{\overline{\nu}_e+\X} and an update of the previously measured
total cross section for \proc{e^{-}p}{\nu_e+\X} \cite{cc94}. The
simplicity of the analysis described in the previous publication is
maintained. Charged current events at 4-momentum
transfer squared \Qsquare\ of order 3000 GeV$^2$ have
the outstanding signature of high transverse momentum hadron systems which
appears unbalanced due to the undetected final state neutrino
(see figure~\ref{fig:ccevt}). This feature makes the visible vector transverse
momentum sum (\Vcc) a simple and efficient discriminant for such events in
getting a clean charged current event sample. Data with \Vcc $>$ 25 GeV were
selected. As the $e^+p$ cross section (a few 10 pb) is less than half
the $e^-p$ cross section, while the background situation remains unaltered,
the background estimation methods previously described had to be somewhat
refined. In particular, a detailed study is
presented on the effect of incoming interacting muons and of
photon-proton interactions. These latter are significant as they are
associated with large cross
sections. The efficiency and correction factors are evaluated directly from
data wherever possible.

The requirement that \Vcc\ be large simplifies the interpretation of the
data in several ways. It enhances the sensitivity to the W propagator
$(Q^2 > 625\ \mbox{GeV}^2)$, it forces Bjorken-$x$ to be
larger than 0.03 and excludes regions in which radiative corrections are
large, ensuring that the prediction of the cross sections is theoretically
reliable.
{}From the measured cross sections their ratio is derived. It is sensitive
to the ratio $D/U$ of the integrated $d$ and $u$ quark distributions as
well as the helicities of the $W^{\pm}$.
%
\section{Experimental conditions}
\subsection{HERA}
This analysis is based on data taken at the $ep$-collider HERA in 1994.
The integrated luminosities for $e^-$ and $e^+$ running were 0.36 and
2.70 pb$^{-1}$ respectively.
The beam energies were 27.6\ GeV\ for electrons and positrons, and 820\GeV\
for protons. HERA was operated with 168 electron and 170 proton bunches.
156 bunches were colliding, separated from each other by 96\ns.
 The width of the interaction region is determined by the length of the
proton bunch and was found to be $\sigma_{\rm zvtx}=10\unit{cm}$.

\subsection{The H1 detector}
A detailed description of the H1 detector and its performance can be
found in \cite{H1}. Below only those aspects which are relevant to the
measurement of the total charged current cross sections are discussed.

The energy of the  hadronic final state  is measured in the highly segmented
liquid argon (LAr) calorimeter  \cite{LAR78}, which cover the polar angular
direction between $4^\circ$  and $153^\circ$ with respect to the proton beam
direction taken to be the $+z$-direction. The LAr calorimenter consists of an
electromagnetic
section with lead absorber and a hadronic section with stainless steel
absorber. The total depth of the electromagnetic part varies with polar angle
between 20 and 30
radiation lengths, whereas the total depth of both calorimeters combined
varies between 4.5 and 8 interaction lengths. The calibration of the
LAr calorimeter was obtained from test beam measurements
using electrons and pions \cite{H1,LAR78,H1PI}. The resolution for pions
was found to be $\sigma(E)/E \approx 0.5/\sqrt(E/\mbox{GeV}) \oplus 0.02$.
The hadron energy scale was verified to $5\%$ from studies of the transverse
momentum balance in neutral current deep inelastic scattering events.
The energy scale for electrons was verified to $3\%$ using the double angle
method\cite{da}.
For trigger purposes the LAr calorimeter is read out in a much coarser
granularity via  FADCs. This information provides a
fast trigger decision based on the energy deposition and on the topology
of the event. The analysis of the FADC spectra allows the determination of the
time at which the event took place with an accuracy of 10~--~30~ns, sufficient
to determine the bunch crossing within which the event occurred.

Charged particles in the polar angle range $15^o - 165^o$ are measured in
the central drift chamber (CJC) and used to determine an event vertex.
The time of occurrence of each event is determined with a precision of 1\ns\
from tracks which cross the CJC sense wire planes.
The CJC is supplemented by several layers of proportional chambers providing
 fast trigger decisions based on tracks.

Both, the LAr calorimeter and the central jet drift chamber are surrounded by a
super-conducting solenoid. The iron  yoke which returns the magnetic flux
is instrumented with streamer tubes and is used as muon detector.
The streamer tubes are read out in digital mode  for trigger purposes and
muon track identification and in analog mode  for energy reconstruction
of particles passing through the LAr calorimeter. The timing of the muon
trigger system is precise enough ($\sim 20$ ns) to assign
events to a unique bunch crossing.

A time-of-flight (ToF)
system built out of two scintillator planes is used to veto
background caused by proton beam-gas and beam-wall interactions upstream of
the detector.

The luminosity system measures the rate of small angle Brems\-strahlung
processes \cite{lumi}.
It consists of two crystal calorimeters at $z=-33$ m and $ z=-103$ m from the
nominal interaction point, the first to detect the scattered electron or
positron {\it (e-tagger),} the other to detect the emitted photon. The
{\it e}-tagger is also used to detect the scattered lepton in $\gamma p$
interactions.
\subsection{Event selection}
Two observables, the {\em scalar} (\Scc) and {\em vector} (\Vcc)
transverse momentum sums, are
used to characterise charged current
events at large values of the 4-momentum transfer square :
\begin{eqnarray*}
\label{eq:SV}
\Scc & \equiv &  \sum_{i}|\ptvec_i| \\
\Vcc & \equiv & |\sum_{i}\ptvec_i| \; .
\end{eqnarray*}
\Scc\ and \Vcc\ correspond to the {\em total transverse energy} and
the {\em missing transverse energy} in each event. The computation of these
quantities is straightforward, requiring only the calo\-rimeter cell energies
(index $i$)
and positions,
and the event vertex position. Charged current events are characterised by
large values of \Scc\ and \Vcc. Note that by definition $\Scc\ge\Vcc$. The
main selection criterion for charged current candidates is based on the
observable \Vcc.

The event selection proceeds in several steps. The hardware trigger condition
for charged current events (\CC) has to be fulfilled. This consists of a
coincidence of the calorimeter \ptmiss -trigger
and the $z$-vertex time signal in anti-coincidence with the ToF-veto signal.
The calorimeter \ptmiss-trigger calculates the vector sum of the transverse
momenta based on the coarsely segmented calorimeter trigger read out and
requires $\ptmiss_{\rm trig} > 6$ GeV. The $z$-vertex time signal identifies
the event time and requires at least
3 out of 4 proportional chamber layers hit.
A fast online reconstruction verifies the hardware trigger decision and
 rejects obvious background from cosmic and
beam halo muons,  beam-gas, and beam-wall interactions.

An event vertex must have been reconstructed from tracks detected in the
central drift chamber. The vertex is required to lie
in the range of $\Delta z$ = $\pm 35$ cm around the nominal interaction point.

The vector sum \ptmiss\ of the transverse momenta, reconstructed from the
LAr calorimeter cells is required to exceed 25 GeV, well above the initial
threshold $V_{\rm trig}$.

In total 1133 candidates satisfied the \CC\ selection criteria.

%
\section{Background}
Two sources of background to the charged current events are
considered: events induced by incoming interacting muons and background from
$ep$ interactions. They are discussed in the two subsequent sections.
\subsection{Muon-induced background}
The charged current selection described above may be satisified by cosmic
ray or proton beam halo events in which the muon interacts with the detector.
These radiative interactions generate electromagnetic showers. A careful
analysis of the topology of these showers, and their timing with respect to
the beam crossing, reveals characteristics that enable the identification
and removal of such events. This analysis is described briefly here and in
more detail in appendix A.

{\bf Halo muons} :
A high flux of muons accompanies the proton beam and some of these traverse the
detector parallel to the beam line. There are two possibilities to fulfill
the \CC\ selection criteria~: \\
(a)     The muon interacts in the LAr calorimeter, while the track of
        the penetrating muon continues over the entire longitudinal
        range of the calorimeter at constant distance from the beam
        line. The secondaries deposit sufficient energy either in the
        electromagnetic or hadronic part of the calorimeter and a
        charged secondary satisfies the vertex criterion. \\
(b)     The halo muon sets the energy trigger, while a simultaneous
        beam related interaction occurred satisfying the vertex requirement.
        Such events are called {\em superimposed event}.

{\bf Cosmic muons} :
Cosmic ray muons penetrate the detector at all angles. Their time of
passage is unrelated to the time of the beam interaction. Again, there are two
ways in which such events may fulfill the \CC\ selection criteria~: \\
(a)     the muon interacts in the detector such that enough energy is
        deposited in the calorimeter and that a secondary charged
        particle accidentally fulfills the vertex requirement \\
(b)     the muon is superimposed over a beam induced interaction (beam
        gas or \GammaP\ ), while the "underlying" event satisfies the
        vertex requirement.

Muon induced background may be identified in two independent ways by applying
topological and timing criteria with the
result of being identified by both, only one, or none of these criteria, as
summarised in table~\ref{tab:mubg}.
\begin{table}
 \begin{center} \begin{tabular}{|l|r|r|}    \hline
 Classification       & {\em Time}-id  & no {\em Time}-id  \\ \hline
 {\em Topo}-id              & 773      & 272         \\
 no {\em Topo}-id           &  18      &  70         \\ \hline
  \end{tabular} \end{center}
 \caption{\sl Classification of the 1133 events according to topology
              ({\em Topo}) and timing ({\em Time}) properties.}
 \label{tab:mubg}
\end{table}
As the topological and timing criteria are independent, the efficiency
of the background identification procedure may also be determined. Removing
events identified as background by either or both of the above criteria
leaves a sample of 70 events. The number of muon associated events remaining
in this sample is $(272\times 18)/773 \approx\ 6 \pm 2$.

In order to check the above procedure, the 88 candidates left untagged by the
topological information are subjected in a visual scan to a quality control.
Of these events 26 events were uniquely verified as being due to incoming
muons in agreement with the number expected from the above analysis of
18 tagged and 6 from the statistical estimate.

The visual scan also revealed 6 obviously misclassified neutral current
events in the \CC\ sample, in which
the electron or positron candidate or the hadron shower point to a crack in
the calorimeter, and 1 event due to noise in the calorimeter. These are
unambiguous background events and are removed from the sample. Furthermore,
there are 2 events containing isolated high \ptnu\ final state
leptons, one with a muon\cite{theevent} and another with an electron.
An obvious source of such events is $W$ production with subsequent leptonic
decay. Events of this type are interesting in their own right and are
excluded from the \CC\ sample.

\begin{figure}[htb]\centering
\mbox{\epsfig{figure=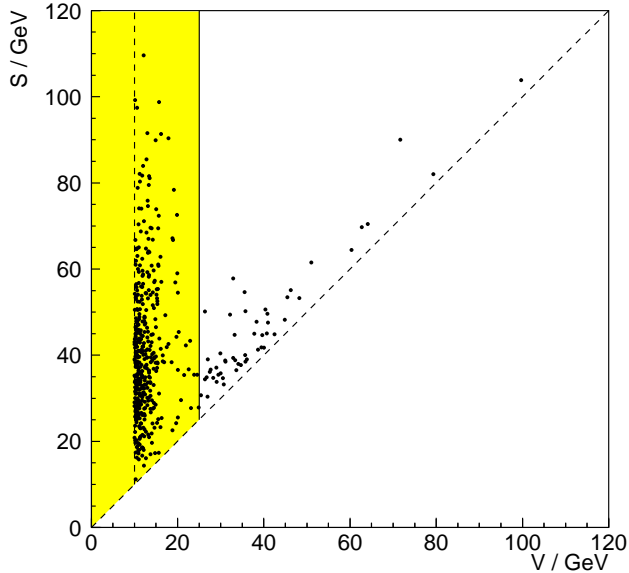,width=8.6cm}}
\caption{\sl Selected events in the \Vcc-\Scc\ plane for \Vcc\ $>$ 10 GeV. The
             \SignalRegion\ is defined by \Vcc $>$ 25 GeV. }
  \label{fig:sv}
\end{figure}

To summarise, after the background subtraction described above, the charged
current sample consists of 53 candidates, 12 within the $e^-$ runs and 41
within the $e^+$ runs. The position of these events in the \Vcc-\Scc\ plane
is shown in fig.~\ref{fig:sv}, the
\SignalRegion\ being that with $\Vcc\ >$ 25 GeV. The expected correlation
between \Vcc\ and \Scc\ for \CC\ events is observed. An outstanding \CC\
event is shown in figure~\ref{fig:ccevt}. The region with $10 < \Vcc\ < 25$
GeV is referred to as {\it control region} in the following.
%
\subsection{$ep$-induced background }

Three sources of $ep$ interaction induced background is considered: beam-gas
or beam-wall interactions; neutral current interactions in which the
electron or positron is not detected
and \GammaP\ interactions. The large transverse momentum imbalance required
($\Vcc\ >$ 25 GeV) eliminates the beam-gas as well as the beam-wall background,
while the other contaminations are strongly suppressed.

\begin{figure}[htb]\centering
\mbox{\epsfig{figure=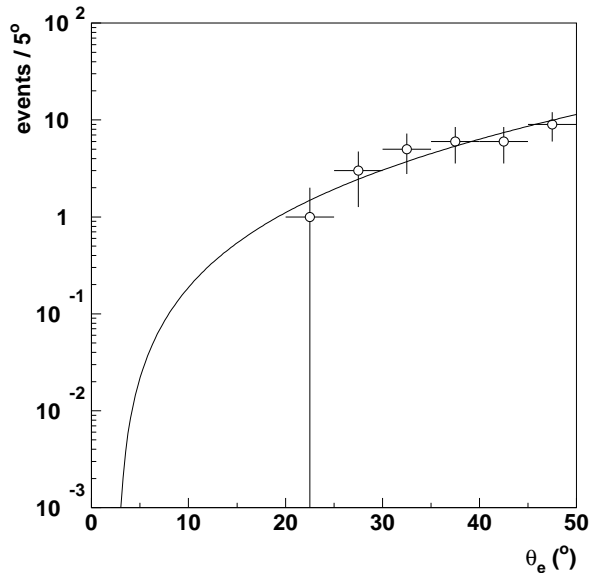,width=8.6cm}}
\caption{\sl Angular distribution of the final state  electron  for a
             neutral current reference sample with \Vcc\ $> 25$ GeV.
             The solid line is the theoretical expectation used for
             extrapolation to small angles. The calorimeter starts at 4$^o$.}
\label{fig:nc}
\end{figure}

A neutral current interaction with large \Qsquare\ may fake a charged
current interaction, if the electron or positron remain undetected. This
may occur if the final state lepton is emitted close to the beam pipe
($\theta_e < 4^o$) or it hits a region of dead material in the detector.
As may be seen from figure~\ref{fig:nc}, which shows the
polar angle distribution of
a reference sample of neutral current events, the former background is
negligible with current data. The latter background was eliminated by
visual inspection.

Neutral current interactions at very small \Qsquare\ may be considered as
being due to the interaction of a quasi real photon with the proton (\GammaP ).
Their cross section is very large. The electron remains usually undetected.
Photon proton interactions with large enough transverse energy show up mainly
as planar two jet configurations which are balanced in transverse momentum,
thus the vast majority of the \GammaP\ events  have \Vcc\ $\approx$ 0 and do
not satisfy the requirement that \Vcc $> 25\,$ GeV.
There are nevertheless two
effects which may cause a \GammaP\ event to have significant transverse
momentum unbalance. These are particle losses in the beam pipe and the
effects of the calorimeter resolution.
Both effects are important as verified by Monte Carlo studies.
Configurations corresponding to \Vcc\ exceeding the cut off value of 25 GeV
require a \GammaP\ interaction in which the participating partons carry
a large proportion of the momentum of the parent particles, resulting also
in a large value of the scalar transverse momentum sum
\Scc. The rate of such events is strongly suppressed.

\begin{figure}[htb]\centering
\mbox{\epsfig{figure=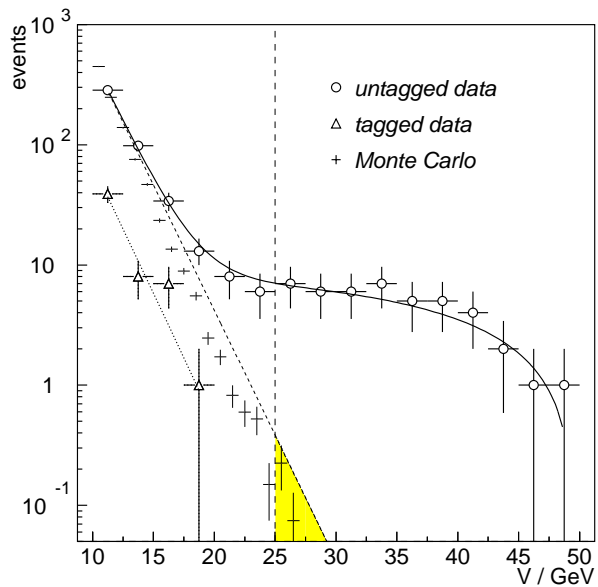,width=8.6cm}}
  \caption{\sl Distribution of all events in the control and signal region
  versus \Vcc\ (the vector transverse momentum sum). The data are marked as
  open circles. The solid line is the result of the   fit described in the
  text. The dashed line shows the \GammaP\ contribution and its
  extrapolation into the signal region. Also shown by open triangles are
  the tagged \GammaP\ data. The \GammaP\ Monte Carlo simulation is represented
  by the small crosses.}
           \label{fig:gammaP}
\end{figure}

In order to assess this background quantitatively the analysis has been
extended to a region with a substantially relaxed \Vcc\ cut of
$10$\GeV, where the detection efficiency for \CC\ events is still
reasonably large.
For this purpose slightly modified selection criteria and less restrictive
trigger requirements were applied.
The additional events
 in this region are
shown in figure~\ref{fig:sv}. Their composition is dominated by \GammaP\
interactions, but contains of course also \CC\ events. Indeed some \GammaP\
events identify themselves unambiguously by the presence of an
isolated final state lepton in the small angle tagging system.
Figure~\ref{fig:gammaP} shows the $V$-distribution of all selected events.
 A steep exponential fall-off is
observed for small values of \Vcc\ in  contrast to the rather flat
distribution at large values of \Vcc\ in the signal region.
 The behaviour for \Vcc\ above 10 GeV
can be well described by the sum of an exponential plus the expected shape of
the \CC\ events. The solid line in figure~\ref{fig:gammaP} is the result of
a fit allowing as free parameters the exponential slope and the normalisations.
The fit yields a slope of $-(0.47 \pm 0.03)$ GeV$^{-1}$.
The extrapolation into the signal region above \Vcc\ = 25 GeV predicts a
background of $0.3 \pm 0.1$ events within the  53 \CC\ candidates.

As a check figure~\ref{fig:gammaP} also shows the predicted behaviour
based on a sample of \GammaP\ events generated using the PYTHIA program
package\cite{mcgp}.
It is normalised to the data at low \Vcc.  The exponential
shape is supported by direct comparison with tagged \GammaP\ events as
seen in figure~\ref{fig:gammaP}.
The exponential behaviour and the numerical value
for the slope agree well with observation corroborating the extrapolation
into the signal region.

As a further check the 14 events in the interval between 20 and 25 GeV
of the control region have been scanned for topologies consisting of two
planar jets. Three such events were found, compatible with the 3.3 \GammaP\
events predicted by the fit.
In conclusion, the background due to \GammaP\ events is  negligible.
%
\section{Corrections}

The efficiencies of and corrections arising from the various selection steps
are summarised in table~\ref{tab:eps}. These were determined using a \NC\
reference sample, termed {\it PseudoCC,}  which was obtained by removing
the electron and requiring for the hadron system the same criteria as for
\CC\ candidates. This procedure works well and does not rely on a
simulation of \CC\ events. Nontheless, the results were checked using a Monte
Carlo simulation and good agreement was found.
\begin{table}[htbp]
\begin{center} \begin{tabular}{| l | c | c |} \hline Analysis step &
Correction factor ($e^-$) & Correction factor ($e^+$) \\
\hline
Selection        & $ 0.97 \pm 0.02 $ & $0.97 \pm 0.02 $ \\
Vertex           & $ 0.86 \pm 0.03 $ & $0.87 \pm 0.03 $ \\
\CC-trigger      & $ 0.90 \pm 0.02 $ & $0.90 \pm 0.02 $ \\
\ptmiss-cut      & $ 0.94 \pm 0.05 $ & $0.91 \pm 0.07 $ \\ \hline
Total $\epsilon$ & $ 0.71 \pm 0.05 $ & $0.69 \pm 0.06 $ \\ \hline
\end{tabular}
\end{center}
\caption[]{\sl Correction factors to the charged current samples.
               } \label{tab:eps}
\end{table}
\begin{figure}[htb]\centering
\mbox{\epsfig{figure=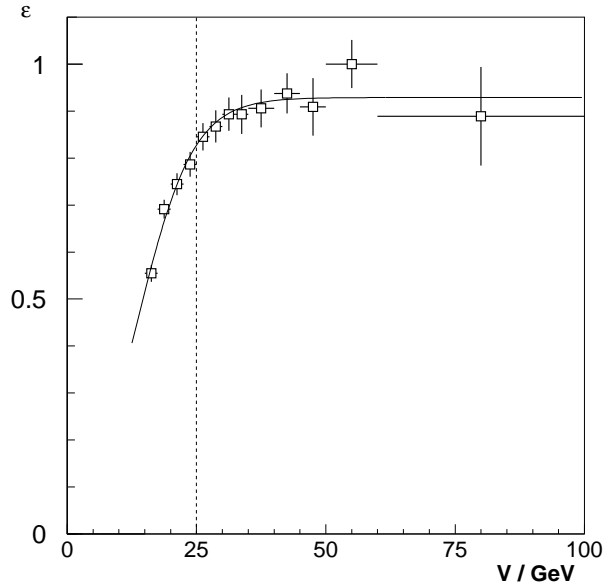,width=8.6cm}}
\caption{\label{fig:trig} \sl The charged current trigger efficiency as a
         function of the vector sum of transverse momenta, \Vcc. The squares
         show the data points and the full line a parametrisation of
         the threshold curve. The dashed line indicates the analysis
         cut.}
\end{figure}
The inefficiency in the \CC\ trigger is mainly due to the coarse granularity
of the calorimeter trigger read out system. The effective threshold is
therefore significantly higher than the nominal one
(cf. fig.\ \ref{fig:trig}). The efficiency for $\ptmiss > 25$ GeV is 0.92. An
additional inefficiency of 0.02 arises from the track requirement in the
trigger.
Together with a small inefficiency from the ToF veto condition, the total
trigger efficiency was determined to 0.90. A systematic error of
0.02 is given by the limited statistics of the {\em PseudoCC} sample.
\begin{figure}[htbp]\centering
\mbox{\epsfig{figure=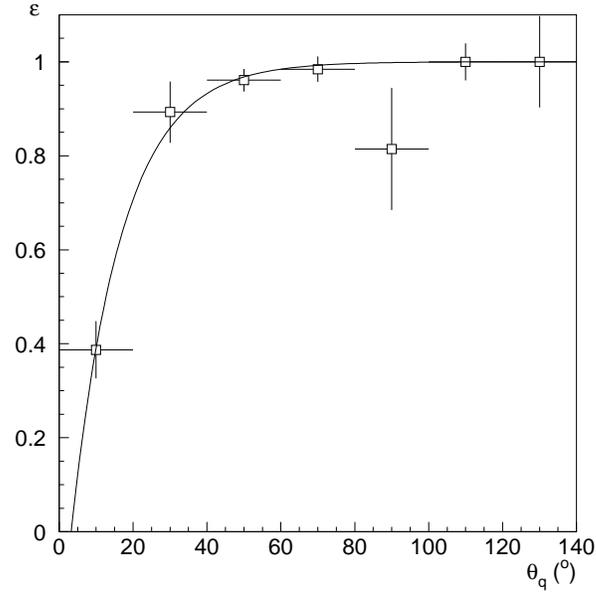,width=8.6cm}}
\caption{\label{fig:vtx} \sl The vertex efficiency as a function of scattering
angle of the hadronic final state. }%
\end{figure}
The limited vertex efficiency is dominantly of geometrical nature. The
 acceptance of the central drift chamber falls steeply in the forward area
below $30^o$ (cf. fig.\ \ref{fig:vtx}). The vertex efficiency for events with
$\ptmiss\ > 25$ GeV is found to be near to 0.9 (see table~\ref{tab:eps}). A
small difference
between the efficiencies for $e^-$ and $e^+$ induced charged current events
is expected, since the angular spectra of their hadronic final
states are different.

The efficiency for the fast online and offline
\begin{figure}[htbp]\centering
\mbox{\epsfig{figure=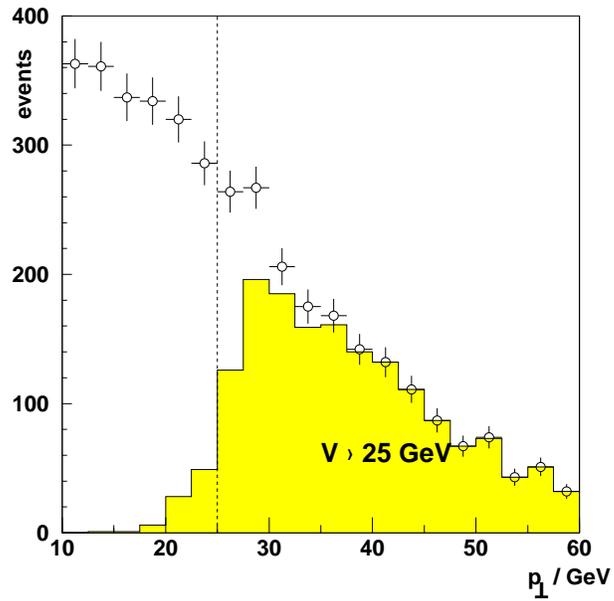,width=8.6cm}}
\caption{\label{fig:mig} \sl Spectrum of the final state neutrino's transverse
     momentum \ptnu\ from a MC simulation of $e^+p$ scattering. The crosses
     show all events, while the shaded histogram shows the distribution for
     those events with $\ptmiss > 25$ GeV. The dashed line represents the
     \ptnu-cut illustrating gain and loss.}
\end{figure}
background rejection chain was as well determined from the {\em PseudoCC}
sample and found to be $0.97 \pm 0.02$.

The reconstructed vector sum of transverse momenta, \ptmiss, is affected
by the apparatus used in the measurement and the details of the method used.
The effect of the \ptmiss-cut is thus experiment dependent and  a correction
must be applied to obtain results corresponding
to the true transverse momentum (\ptnu) of the final state neutrino.
Figure \ref{fig:mig} illustrates how the \ptmiss-cut
correction accounts for gains and losses due to resolution, radiative
effects, and particles escaping detection.
The calculation of the correction factor relies on MC simulation and yields
0.94 and 0.91 for the $e^-$ and $e^+$ running, respectively.
The migration of events depends strongly on the energy scale of the
LAr calorimeter. For the estimated $\pm 5\%$ hadron
energy scale error the migration efficiency
changes by 0.05 and 0.07 for $e^-$ and $e^+$. These uncertainties
are accounted for in the systematic error. Theoretical uncertainties in the
hadronisation model used in the  simulation yield an additional error of
0.01.

\section{Results and conclusion}

The charged current cross sections for $\ptnu>$25 GeV are then
calculated according to:
\begin{eqnarray*}
  \sigma = \frac{N}{\La \epsilon}.   \nonumber
\end{eqnarray*}
The data samples used correspond to an integrated luminosity of
  $   \La  = 2.70 \pm 0.05 \, {\rm pb}^{-1}$ for $e^+$ and
  $   \La  = 0.36 \pm 0.01 \, {\rm pb}^{-1}$ for $e^-$.
The statistical errors of these measurements are negligible.
The systematic errors arise primarily from uncertainties as to the
acceptance of the luminosity system under varying beam
conditions. Considerable progress has been made in understanding these
uncertainties since 1993. The observed numbers of 12 $e^-$ and 41 $e^+$ induced
charged current events, with the corrections listed in table~\ref{tab:eps},
give cross sections of:
\begin{eqnarray*}
 & \sigma&(e^-p|\ptnu>25\GeV)  =  46.6 \pm 13.5 \pm 3.5  \, {\rm pb} \\
 & \sigma&(e^+p|\ptnu>25\GeV)  =  21.9 \pm\ 3.4 \pm 2.0  \, {\rm pb}
\end{eqnarray*}
where the first error is statistical and the second error includes all
systematic effects added in quadrature. The systematic uncertainties
of the two results are strongly correlated.

\begin{figure}[htbp]\centering
 \mbox{\epsfig{figure=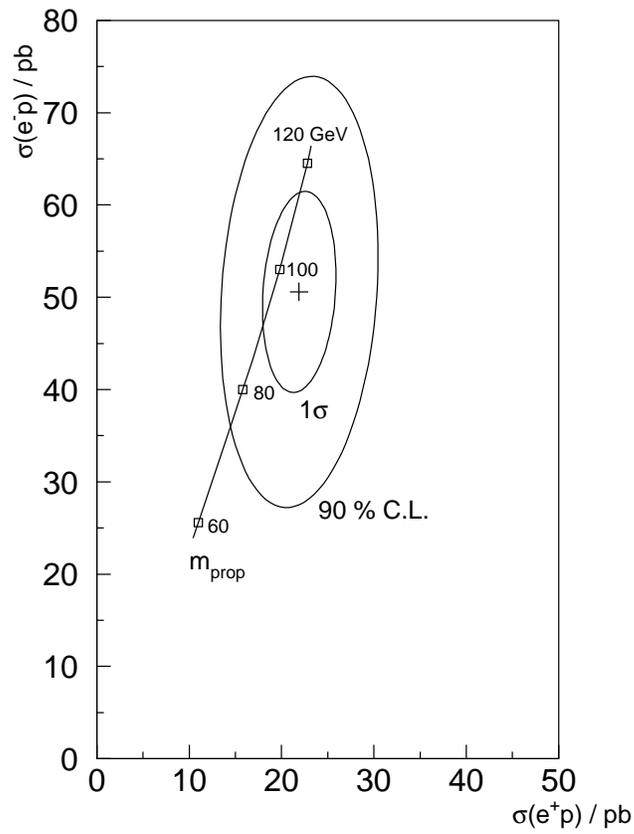,width=8.6cm}}
   \caption{\sl The measured $e^-p$ and $e^+p$ cross sections compared
            with the predicted cross section as a function of the W
            propagator mass ($m_{\rm prop}$). The two ellipses correspond to
            the 39.4 \% (1$\sigma$) and 90~\% CL contours.}
            \label{fig:sigsig}
\end{figure}

The measured total $e^- p$ cross section agrees well with value previously
published by the H1 Collaboration\cite{cc94} as well as with
the measurement of the ZEUS Collaboration\cite{zeus}. Combination of the
earlier H1 result with that discussed here, taking into account common
systematics (the correlation coefficient is 0.70) and the increased electron
beam energy gives:
\begin{eqnarray*}
  \sigma(e^-p|\ptnu>25\GeV) = 50.6 \pm 10.9 \, {\rm pb}
\end{eqnarray*}
where the error includes both statistical and systematic uncertainties.
The correlation coefficient of the systematic errors of the combined $e^-p$
and the $e^+p$ measurements is 0.16, which leads to a slight tilting of the
error ellipse in the $\sigma(e^-)-\sigma(e^+)$ plane shown in
fig.~\ref{fig:sigsig}. This figure also shows the theoretical expectation as a
function of the
$W$ propagator mass, calculated using the HERACLES\cite{theory} package
with the inclusion of electroweak radiative effects \cite{hs}. The sensitivity
of both the $e^-p$ and $e^+p$ cross sections to the
mass value assumed for the $W$-propagator is clearly visible. The best
value of the propagator mass inferred from this figure agrees within 90~\%
C.L. with the  known resonance mass of 80.2\GeV~\cite{pdg}, while its
uncertainty is about 15 GeV.

The charged current event sample includes 2 events with isolated
photons of transverse momentum larger than 2 GeV,  consistent with the theore%
tical expectation.

The ratio $R_e$ of the two measured charged current cross sections is
\begin{eqnarray*}
  \frac{\sigma(e^+p|\ptnu>25\GeV)}{\sigma(e^-p|\ptnu>25\GeV)}
       = 0.43 \pm 0.11,
\end{eqnarray*}
where the common systematic uncertainties drop out.

The quark-parton model explains qualitatively the value of the cross
section ratio $R_e$ in terms of the relevant $e^{\pm}$ quark and
antiquark subprocesses mediated by the weak vector bosons $W^{\pm}$. A good
approximation is :
\begin{eqnarray*}
R_e \approx \frac{\sigma(e^+d)+\sigma(e^+\overline{u})}{\sigma(e^-u)}
     = \frac{D}{U}(a_1 + a_0\frac{\overline{U}}{D})
\end{eqnarray*}
where  $D/U$ and $\overline{U}/D$ are the ratios of the corresponding
integrated parton distributions, whereas $a_1= \langle (1-y)^2 \rangle $
($y$ being the Bjorken scaling variable) and $a_0= \langle 1 \rangle $ account
for the $W$ helicities averaged over the appropriate parton distributions
\cite{vb}.
While the sea in the proton contributes little to $e^-p$ scattering, its
contribution to $e^+p$ scattering is of the same size as that of the one of
the valence $d$-quark. Although the $W$ propagator strongly affects each cross
section separately, the effect on cross section ratios is much reduced.

Finally, the measured ratio $R_e$ can be compared with the analogous
quantity, the ratio $R_{\nu}$, measured previously in fixed
target neutrino experiments \cite{hp} at much lower energies :
\begin{eqnarray*}
  \frac{\sigma(\proc{\nu_{\mu} p}{\mu^-+\X})}
       {\sigma(\proc{\overline{\nu}_{\mu}p}{\mu^++\X})}
       = 0.97 \pm 0.04.
\end{eqnarray*}
%
Both $R_e$ and $R_{\nu}$ are the ratio of a $W^+$ and a $W^-$ induced
interaction on a proton target, however at vastly different energies.
With the approximation of $R_{\nu}$ by
$D/U \cdot (a_1 + a_0\cdot \overline{D}/U)^{-1}$, the numerical difference
between $R_e$ and $R_{\nu}$ can be understood as the interplay of the $W$
helicities and the quark flavors. The large difference in the interaction
energies modifies the averages over the angular distributions ($a_1$ and
$a_0$) and influences considerably the importance of the sea in the proton.

The quantitative comparison of the two ratios is done without approximations.
$R_e$ is first extrapolated to \ptnu = 0 implying the use of the
structure functions also at values of
Bjorken-$x$ below 0.03. The MRS-H \cite{mrs} parton distributions are used.
They are based on HERA data as well as low energy fixed target measurements.
They also agree well with measurements at the TEVATRON\cite{cdf} which are
sensitive to the $u$ and $d$ quark distributions. The extrapolation in \ptnu
down to 0 emphasises the contribution of the sea and increases the $D/U$
ratio by about 20~\%. The ratio $R_e$ after extrapolation to \ptnu = 0 becomes
$0.59 \pm 0.15$ and can be converted into $R_{\nu}$ by calculating the effect
of replacing the $e$ by a $\nu$ and by extrapolating from HERA energies
corresponding to about 48 TeV, if seen as fixed target experiment, down to
SPS energies of the order of 100 GeV.
The ratio $R_e$ after applying all relevant conversion factors
becomes  $1.15 \pm 0.30$ in good agreement with $R_{\nu}$= 0.97 $\pm$ 0.04
measured in neutrino experiments.

\vspace{1cm}
\par\noindent
{\bf Acknowledgement}

\noindent
We are grateful to the HERA machine group whose outstanding efforts
made this experiment possible. We appreciate the immense effort of the
engineers and technicians who constructed and maintained the detector.
We thank the funding agencies for financial support. We acknowledge the
support of the DESY technical staff. We also wish to thank the DESY
directorate for the hospitality extended to the non-DESY members of the
collaboration.
We thank H. Spiesberger for his help in the comparison
with theory.
\vspace{1cm}
%
\section*{Appendix : Incoming muon background}
%
A detailed account is presented on the characteristics and treatment of
the background originating from incoming muons.
\subsection*{Characteristics}
Incoming muons appear either as halo or as cosmic muons with well defined
and unique topological and timing properties which may serve to a twofold
classification.

{}From the topological point of view the 1133 events are attributed to three
classes (see table~\ref{tab:halcos})~:

The {\bf Halo} class consists of 509 events identified by the halo finder.
The principle of halo finders is to check for an energy distribution
in the LAr calorimeter which is very localised in the $r-\phi$-plane
while having large spread in the $z$-direction thus being typical
for a particle traversing the detector parallel to the beam. This
information is combined with a requirement of muon tracks or energy
deposits in the instrumented iron yoke.

The {\bf Cosmic} class consists of 536 events identified by the cosmic
but not the halo finder.
Cosmic finders use  moments of energy distributions in the
calorimeters, local properties (like direction and shape) of calorimeter
showers as well as tracks and energy deposits in the muon detectors to
identify cosmic muons which deposit large amounts of energy in the
detector.

The {\bf CC} class consisting of 88 charged current candidates contains
  the genuine \CC\ events together with a small background of incoming
  muons due to the inefficiencies of the finders.

The \CC\ class appears as the rest class of events neither assigned to
the halo nor the cosmic class. The three classes are disjoint.
\begin{figure}[htbp]
\begin{minipage}{0.49\textwidth}
\mbox{\epsfig{figure=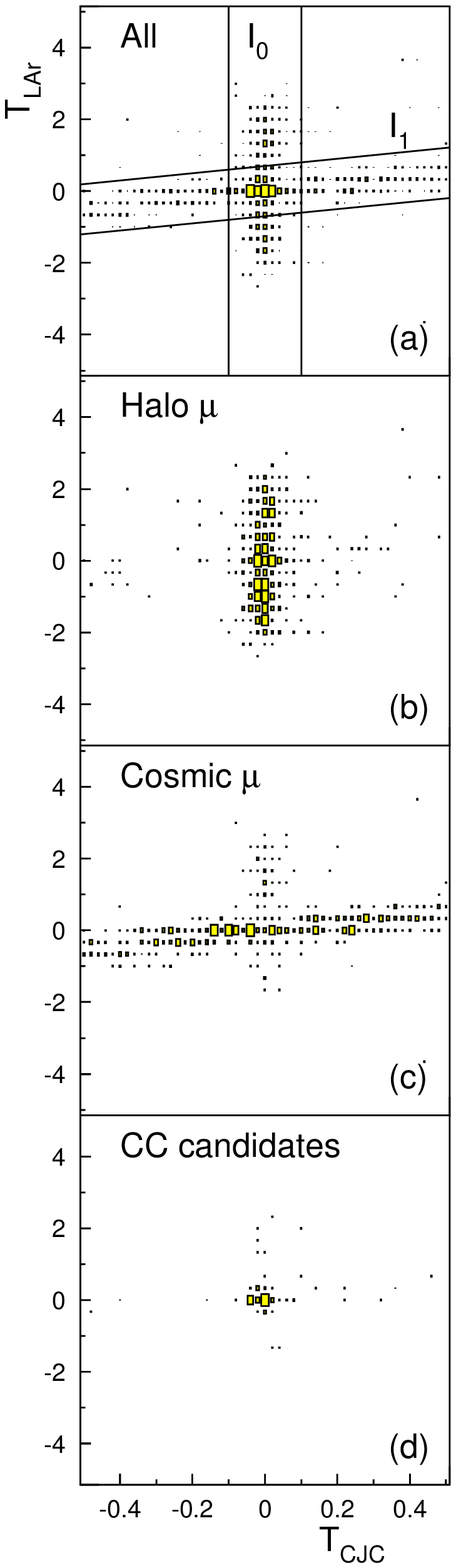,height=0.8\textheight}}
\caption{\label{fig:time}
               \sl The time structure of all events (a), halo events(b),
                   cosmic muon events (c) and \CC\ candidates (d) is
                   shown as a function of the CJC and the LAr timing. All times
                   are given in units of 1 bunch crossing (96 ns). The two
                   bands $I_0$ and $I_1$ are explained in the text. }
\end{minipage}
\hfill
\begin{minipage}{0.49\textwidth}
\mbox{\epsfig{figure=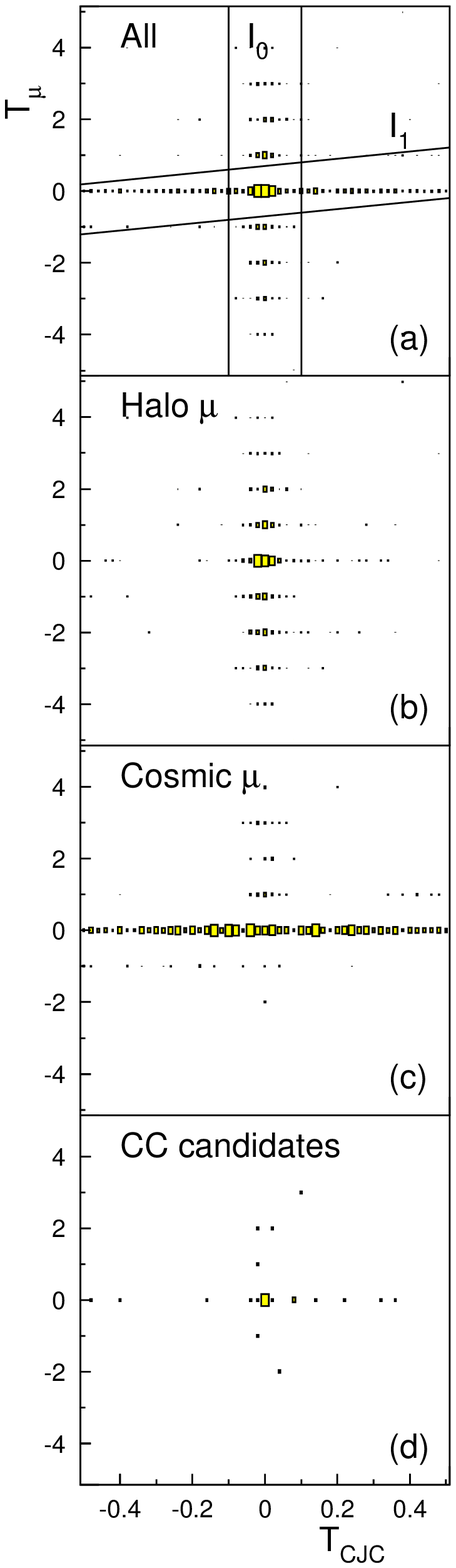,height=0.8\textheight}}
\caption{\label{fig:timeMU} \sl
                   The time structure of all events with a signal in the
                   muon detector (a), halo events(b),
                   cosmic muon events (c) and \CC\ candidates (d) is
                   shown as a function of the CJC and the $\mu$ timing. All
                   times are given in units of 1 bunch crossing (96 ns). The
                   two bands $I_0$ and $I_1$ are explained in the text. }

\end{minipage}
\end{figure}
The scatter plot (figure~\ref{fig:time} a) of the 1133 events in
terms of the timing of the Jet chamber and the LAr calorimeter shows
two prominent time bands. To be specific, two time bands  $I_0$ and
$I_1$  are defined by $T_{CJC} = 0.0 \pm 0.1$ in units of the bunch
crossing time and by $|T_{CJC}-T_{LAr}| <$ 0.7 respectively. The
intrinsic timing accuracy of the Jet chamber is better than 1 ns. The
larger band of $\pm 0.1$ ensures independence of assumptions on
the origin and direction of the particles. The upper limit of $0.7$
comfortably takes into account the variations of the LAr timing
resolution over the entire calorimeter. The two bands distinguish
events according to whether they occur correlated or uncorrelated with
the bunch crossings. Thus, each event can then be assigned in an independent
way to exactly one of four time windows~:

 342 {\bf prompt} events (P) in the intersect of $I_0$ and $I_1$

 414 {\bf uncorrelated} events (U), i.e. not prompt events in
      the band $I_1$

 344 {\bf superimposed} events (S), i.e. not prompt events in
      the band $I_0$

 33  {\bf non assigned} events (N) belonging neither to $I_0$ nor to $I_1$

The U, S and N classes contain only background events, whereas  genuine
\CC\ events belong to the P class. All classes are again disjoint by
\begin{table}
 \begin{center} \begin{tabular}{|l|r|r|r|r|r|}    \hline
 Classification       & All   &  P   &  U   &   S   &  N  \\ \hline
 All                  & 1133  & 342  & 414  & 344   &  33 \\ \hline
 Halo                 & 509   & 159  &  38  & 287   &  25 \\
 Cosmics              & 536   & 113  & 366  &  50   &   7 \\
 CC candidates        &  88   &  70  &  10  &   7   &   1 \\ \hline
 \end{tabular} \end{center}
 \caption{\sl Event classification according to topology and
              CJC and LAr timing. The abbreviations P,U,S,N stand for
              prompt, uncorrelated, superimposed, non assigned events.}
 \label{tab:halcos}
\end{table}
construction. \\

The combination of the two classification schemes leads to 12 disjoint
subclasses as listed in table~\ref{tab:halcos} and shown in
figure~\ref{fig:time}.
\begin{table*}
 \begin{center} \begin{tabular}{|l|r|r|r|r|r|r|r|}    \hline
 Classification  & all & $\mu$ corr(seen) & P & U & S & N \\ \hline
 halo            & 509 & 509 (409)  & 135.7 (109) &
                         34.8(\ 28)  & 304.9 (245) & 33.6 (27) \\
 cosmics         & 536 & 536 (508)  & 112.9 (107) &
                         364.0(345) & 48.5  (\ 46) & 10.6 (10) \\
 CC candidates   &  88 & 26.3 (\ 23) & 11.4  (\ 10) &
                         8.2 (\ \ 7) & 5.7  (\ \ 5) &  1.1 (\ 1) \\ \hline
 \end{tabular} \end{center}
 \caption{\sl Event classification according to topology and CJC
              and $\mu$ timing corrected(uncorrected) for $\mu$
              detection efficiency.}
 \label{tab:hcmu}
\end{table*}
\subsection*{Determination of the background due to incoming muons}
Table~\ref{tab:halcos} shows that the topological and the timing
information alone efficiently reduce the background. It is
their simultaneous application which allows to disentangle the
incoming muon background quantitatively.

Halo events are expected to populate the $I_0$ band. This is veri%
fied by figure~\ref{fig:time} b. Similarly, cosmic muons occur at the
same time as recorded by both subdetectors, but unrelated with
the bunch crossing time, i.e. they are expected to populate the  $I_1$
band, as borne out by figure~\ref{fig:time} c.

The observed number of prompt cosmic muon events matches well with the
expected number calculated from the side bands. This holds for the
non superimposed halos as well. It is
interesting to note that figure~\ref{fig:time} d is dominated by
prompt events, the expected genuine \CC\ events, but still contaminated
with both halo and cosmic muon events, as inferred from the observed 18
nonprompt events. In fact, the observed number of superimposed and
uncorrelated events in the three topological classes (see
table~\ref{tab:halcos}) allows the calculation of the inefficiency of
the halo and cosmic filters yielding $1-\epsilon = 0.02 \pm 0.01$
and $0.03 \pm 0.01$ respectively. The 70 prompt \CC\ candidates
have therefore a residual background of $6 \pm 2$ events due to
incoming muons.

The timing of the muon trigger system allows an independent check of the
above. Since \CC\ events do not have muons (apart from
events containing decay muons), the use of the muon timing
checks directly the background due to incoming muons. Although the
precise muon timing of 20 ns is truncated to the corresponding bunch
crossing time and is not available for all incoming muons due to the
inefficiency of the muon detection, it spans a wide range of
bunch crossings (see figure~\ref{fig:time} b).
Figure~\ref{fig:time} and table~\ref{tab:hcmu} repeat the
corresponding figure and table with the muon trigger timing
replacing the LAr timing.
The numbers in table~\ref{tab:hcmu} include a correction for
muon detection efficiency ( 80 \% for halos and
95 \% for cosmics) are very similar to the corresponding numbers in
table~\ref{tab:halcos} except of course for the \CC\ class expected
to contain primarily background. The inefficiencies of the filters
estimated from table~\ref{tab:hcmu} agree with the numbers derived
from table~\ref{tab:halcos} and lead to an estimated background of
$5\pm 2$ events among the 10 observed prompt events which becomes 11.4
events after efficiency correction.
\subsection*{Visual scan}
The purpose of the visual scan, as mentioned in section 3, was to check the
quality of the 88 events in the \CC\ class and to identify and remove
residual background events from the sample.

The 18 non-prompt events of table~\ref{tab:halcos} were indeed found to
be due to incoming muons.  The remaining 17 prompt background events
contain 8 incoming muons (7 cosmic muons and one halo muon) also be
compared with the $6\pm 2$ events calculated above.

The scan results can also be confronted with the 23 events among the 88
events for which a muon signal is available (table~\ref{tab:hcmu}). The 13
non-prompt events in this sample were identified in the scan as
10 cosmic and 3 halo muons.

The 5 observed prompt incoming muons (4 cosmic and one halo muon) are in
agreement with the statistical estimate of $5\pm 2$. Two of the prompt events
are identified as \CC\ events which happen to leave a signal in the muon
detector either as a result of a single penetrating track or by leakage of
the hadronic
shower. The full prompt muon signal is then quantitatively explained after
accounting for two observed \NC\ events and  the peculiar muon
event~\cite{theevent}.

%

\end{document}